\documentclass[12pt,prd,amsmath,amssymb]{article}
\usepackage{graphicx}
\usepackage[percent]{overpic}
\usepackage{hyperref}

\newcommand\pubnumber{SNSN-xxx-xx}
\newcommand\pubdate{\today}

\def\napoli{University of Illinois at Urbana-Champaign, 61801- Urbana, IL, USA,\\
for the CDF Collaboration.}

\def\Title#1{\begin{center} {\Large #1 } \end{center}}
\def\Author#1{\begin{center}{ \sc #1} \end{center}}
\def\Address#1{\begin{center}{ \it #1} \end{center}}

\newcommand\pubblock{\rightline{\begin{tabular}{l} \pubnumber\\
         \pubdate  \end{tabular}}}
\newenvironment{Abstract}{\begin{quotation}  }{\end{quotation}}
\newenvironment{Presented}{\begin{quotation} \begin{center} 
             PRESENTED AT\end{center}\bigskip 
      \begin{center}\begin{large}}{\end{large}\end{center} \end{quotation}}

\def\beq{\begin{equation}}
\def\eeq#1{\label{#1}\end{equation}}
\def\eeqn{\end{equation}}

\def\beqa{\begin{eqnarray}}
\def\eeqa#1{\label{#1}\end{eqnarray}}
\def\eeqan{\end{eqnarray}}
\def\CR{\nonumber \\ }

\let\bar=\overbar

\def\vev#1{\langle #1 \rangle}

\def\O{{\cal O}}

\def\Dslash{\not{\hbox{\kern-4pt $D$}}}
\def\dslash{\not{\hbox{\kern-2pt $\del$}}}

\def\msb{{\bar{\ssstyle M \kern -1pt S}}}

\newcommand{\CP}{{\ensuremath{C\!P}}}
\newcommand{\Acp}{\ensuremath{\mathcal{A}_\CP}}
\newcommand{\stat}{\ensuremath{\mathrm{~(stat)}}}
\newcommand{\syst}{\ensuremath{\mathrm{~(syst)}}}
\newcommand{\Dbar}{\ensuremath{\overline{D}{}}}

\newcommand{\tev}{\ensuremath{\mathrm{Te\kern -0.1em V}}}
\newcommand{\gev}{\ensuremath{\mathrm{Ge\kern -0.1em V}}}	
\newcommand{\mev}{\ensuremath{\mathrm{Me\kern -0.1em V}}}	
\newcommand{\kev}{\ensuremath{\mathrm{ke\kern -0.1em V}}}	
\newcommand{\massgev}{\mbox{\gev/$c^2$}}			
\newcommand{\massmev}{\mbox{\mev/$c^2$}}			
\newcommand{\pmev}{\mbox{\mev/$c$}}				

\begin{document}
\begin{titlepage}
\pubblock

\vfill
\Title{Search for CP violation in $D^0 \to h^+h^-$ decays at CDF\\ \vspace*{2.0cm}}

\vfill
\Author{ Sabato Leo}
\Address{\napoli}
\vfill
\begin{Abstract}
I report on a measurement of \CP-violating asymmetries ($A_{\Gamma}$) between effective lifetimes of $D^0$ or $\Dbar^0$ in fully reconstructed \mbox{$D^0\to K^+ K^-$} and \mbox{$D^0\to \pi^+\pi^-$} decays collected in $p\bar{p}$ collisions by the Collider Detector at Fermilab experiment. The full CDF data set corresponding to $9.7$~fb$^{-1}$ of integrated luminosity is used. The flavor of the charm meson at production is determined by exploiting the strong-interaction decay $D^{*+} \to D^0 \pi^+$, while the contamination from mesons originated in $b$-hadron decays is evaluated and subtracted from the sample. Signal yields as functions of the observed decay-time distributions are extracted from maximum likelihood fits and used to measure the asymmetries. The results, $A_\Gamma (K^+K^-) = \bigl(-1.9 \pm 1.5 \stat \pm 0.4 \syst\bigr)\times10^{-3}$ and $A_\Gamma (\pi^+\pi^-)= \bigl(-0.1 \pm 1.8 \stat \pm 0.3 \syst\bigr)\times10^{-3}$, and their combination, $A_\Gamma = \bigl(-1.2 \pm 1.2)\times10^{-3}$, are consistent with the SM predictions and other experimental determinations. 

\end{Abstract}
\vfill
\begin{Presented}
 8th International Workshop on the CKM Unitarity Triangle (CKM 2014),Vienna, Austria, September 8-12, 2014.
\end{Presented}
\vfill
\end{titlepage}
\def\thefootnote{\fnsymbol{footnote}}
\setcounter{footnote}{0}

\section{Introduction}
In charm transitions, the standard model (SM) predicts \CP-violating effects not exceeding  \O$(10^{-2})$~\cite{theory}. Indeed, no \CP-violating effects have been firmly established yet in charm dynamics~\cite{hfag}.\par
The decay-time-dependent rate asymmetries of charm mesons two body hadron decays ($D \to h^+h^-$, $h=K,\pi$),
\begin{equation}\label{eq:acp}
\Acp(t) = \frac{d\Gamma(D^0\to h^+h^-)/dt - d\Gamma(\Dbar^0\to h^+h^-)/dt}{d\Gamma(D^0\to h^+h^-)/dt+d\Gamma(\Dbar^0\to h^+h^-)/dt},
\end{equation}
probe non-SM physics contributions in the {\it oscillation} and {\it penguin} transition amplitudes which could be affected by non-SM contributions that enhance the magnitude of the observed \CP\ violation w.r.t.~SM expectation. The asymmetry $\Acp(t)$ includes both direct and indirect \CP\ violation effects. 
The slow oscillations rate~\cite{hfag} of charm mesons allows approximating Eq.\ (\ref{eq:acp}) as \cite{AcpCDF},

\beq
\Acp(t) \approx \Acp^{\rm{dir}}(h^+h^-) - \frac{\vev{t}}{\tau}\ A_\Gamma(h^+h^-) \mbox{with} A_{\Gamma} = \frac{\hat{\tau}(\Dbar^0\to h^+h^-)-\hat{\tau}(D^0\to h^+h^-)}{\hat{\tau}(\Dbar^0\to h^+h^-)+\hat{\tau}(D^0\to h^+h^-)}
\eeq{eq:acp3}
where $\vev{t}$ is the sample mean of decay time, $\tau$ is the \CP-averaged $D$ lifetime~\cite{hfag}, $\Acp^{\rm{dir}}$ is related to direct \CP\ violation, and $A_\Gamma$ is the asymmetry between the {\it effective} lifetimes $\hat{\tau}$ of $D^0$ and $\Dbar^0$ and is mostly due to indirect \CP\ violation. 
Recent measurements of $A_{\Gamma}$~\cite{BfactoriesAgamma} showed consistency with \CP\ symmetry with $\mathcal{O}(10^{-3})$ uncertainties. However, additional determinations with comparable precision may improve the knowledge of \CP\ violation in the charm sector. 
In this note I report a measurement of $A_{\Gamma}$ using  fully reconstructed \mbox{$D^0\to K^+ K^-$} and \mbox{$D^0\to \pi^+\pi^-$} decays collected in $p\bar{p}$ collisions by the Collider Detector at Fermilab experiment. The full CDF data set corresponding to $9.7$~fb$^{-1}$ of integrated luminosity is used.

\section{Selection and reconstruction}

\begin{figure}[t]
\centering
\begin{overpic}[width=0.45\textwidth]{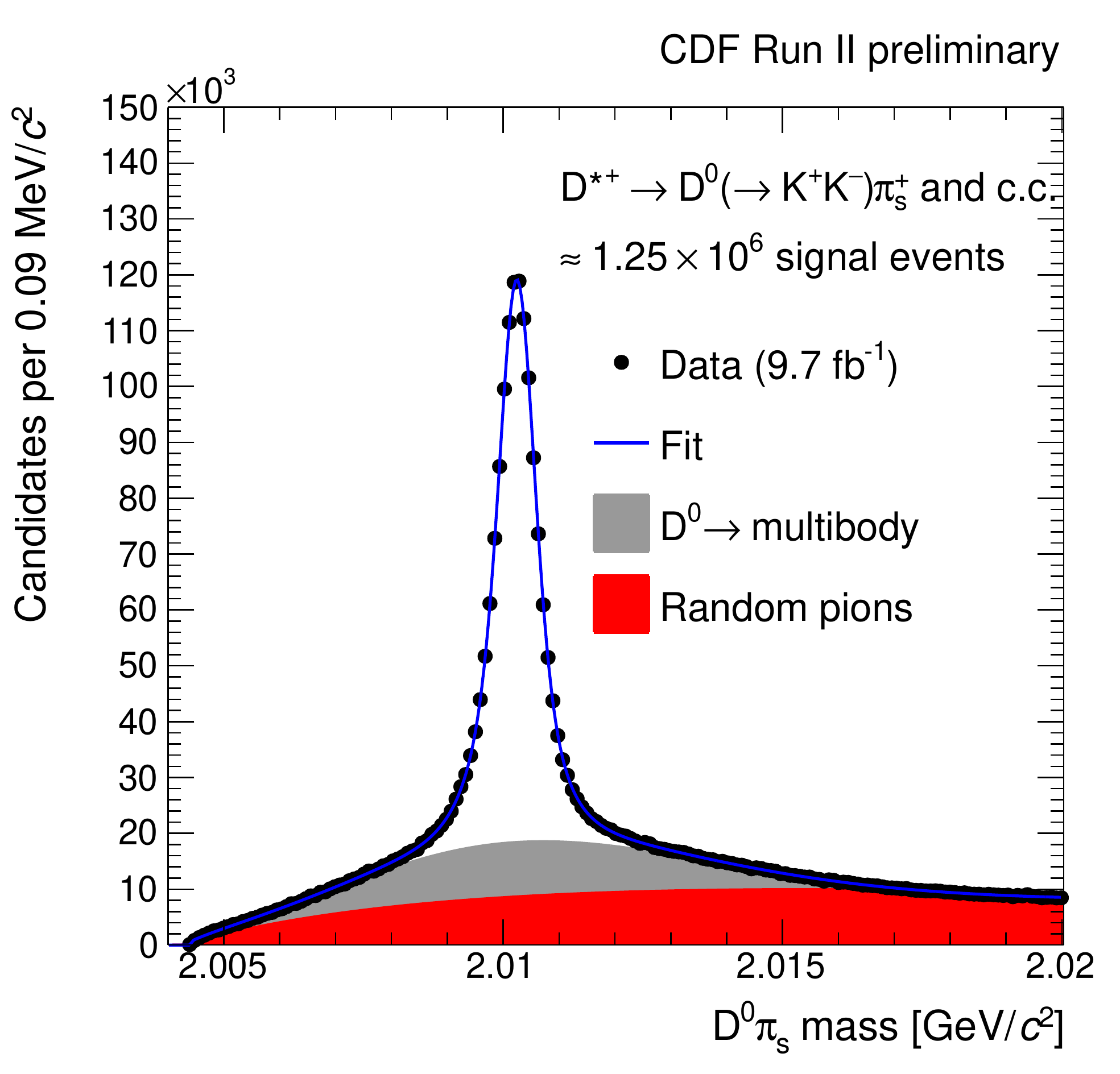}
\put(25,25){(a)}
\end{overpic}\hfil
\begin{overpic}[width=0.45\textwidth]{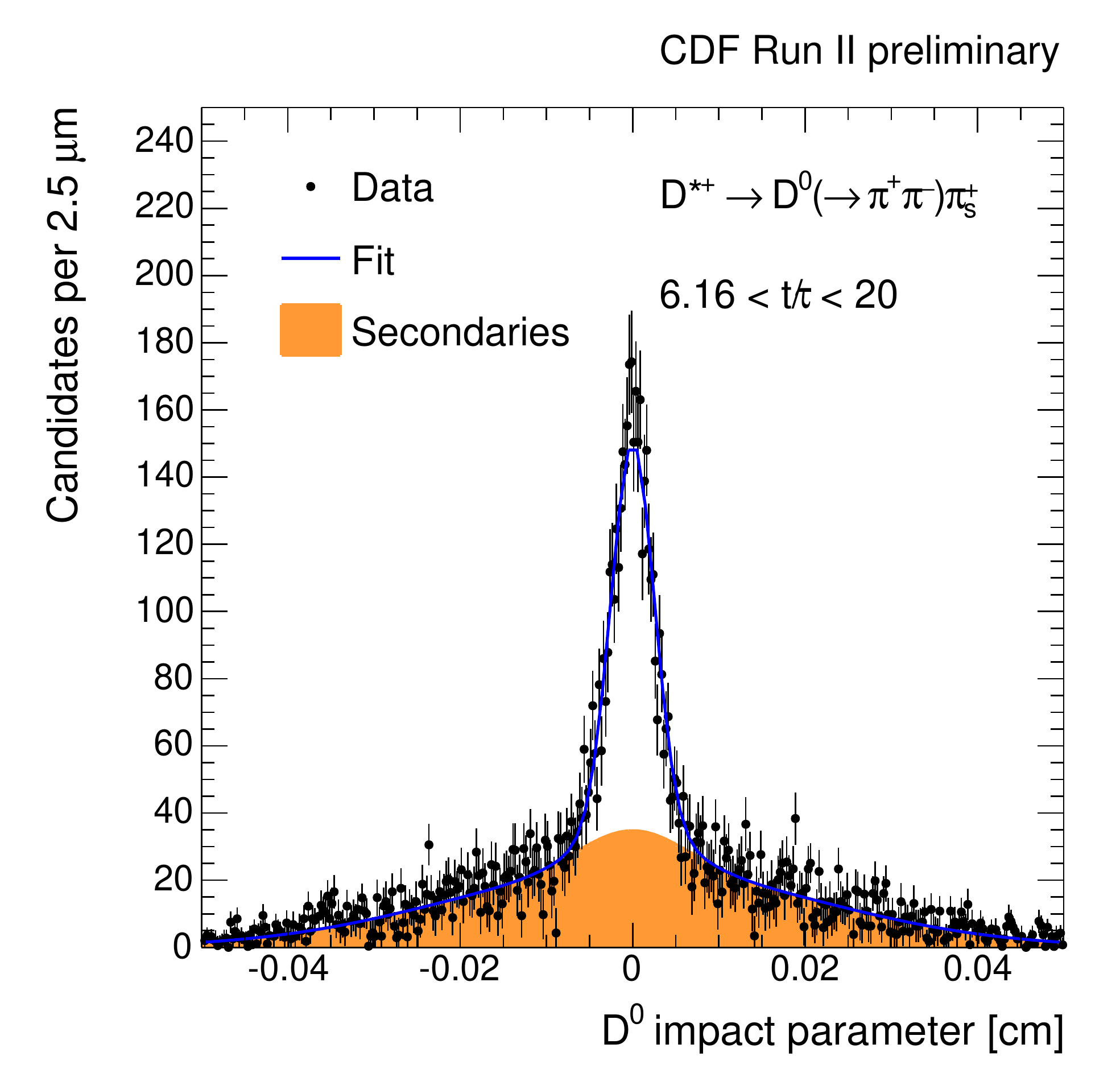}
\put(25,25){(b)}
\end{overpic}\\
\caption{Distributions of $D\pi^{\pm}$ mass with fit results overlaid for the $D \to K^+K^-$ sample (a). Distributions of $D^0$ impact parameter with fit results overlaid for background-subtracted $D^{*+} \to D^0 (\to \pi^+\pi^-)\pi_s^+$ decays restricted to the decay-time bin $6.16 < t/\tau < 20$ (b).}\label{fig:mass}
\end{figure}

Online data selection is based on pairs of charged particles displaced from the $p\bar{p}$ collision point. Offline, a $D$ candidate is reconstructed using two oppositely charged tracks fit to a common decay vertex. A charged particle with $p_T>400~\pmev$ is associated with each $D$ candidate to form $D^{*\pm}$ candidates. 
Constraining the $D^{*\pm}$ decay vertex to lie on the beam-line results in a 25\% improvement in $D^{*\pm}$ mass resolution w.r.t Ref.~\cite{AcpCDF}.  Ref.~\cite{AcpCDF} details the offline selection. The $h^+h^-$ mass of selected candidates is required to be within about 24 \massmev\ of the known $D^0$ mass, $m_{D^0}$~\cite{hfag}, to separate $D \to K^+K^-$ and $D \to \pi^+\pi^-$ samples.
Final selected samples contain $6.1\times 10^5$ $D^0 \to K^+K^-$,  $6.3\times 10^5$ $\overline{D}^0 \to K^+K^-$, $2.9\times 10^5$ $D^0 \to \pi^+\pi^-$, and $3.0\times 10^5$ $\overline{D}^0 \to \pi^+\pi^-$ signal events.The main backgrounds are real $D^0$ decays associated with random pions or random combinations of three tracks (combinatorics) for the $\pi^+\pi^-$ sample, while the $K^+K^-$ sample is also polluted by misreconstructed multibody charm meson decays (i.e. $D^0 \to h^-\pi^+ \pi^0$ and $D^0 \to h^- \ell^+ \nu_{\ell}$, where $\ell$ is a muon or an electron), Figure.~\ref{fig:mass}. 

\section{Determination of the asymmetry}

\begin{figure}[t]
\centering
\begin{overpic}[width=0.45\textwidth]{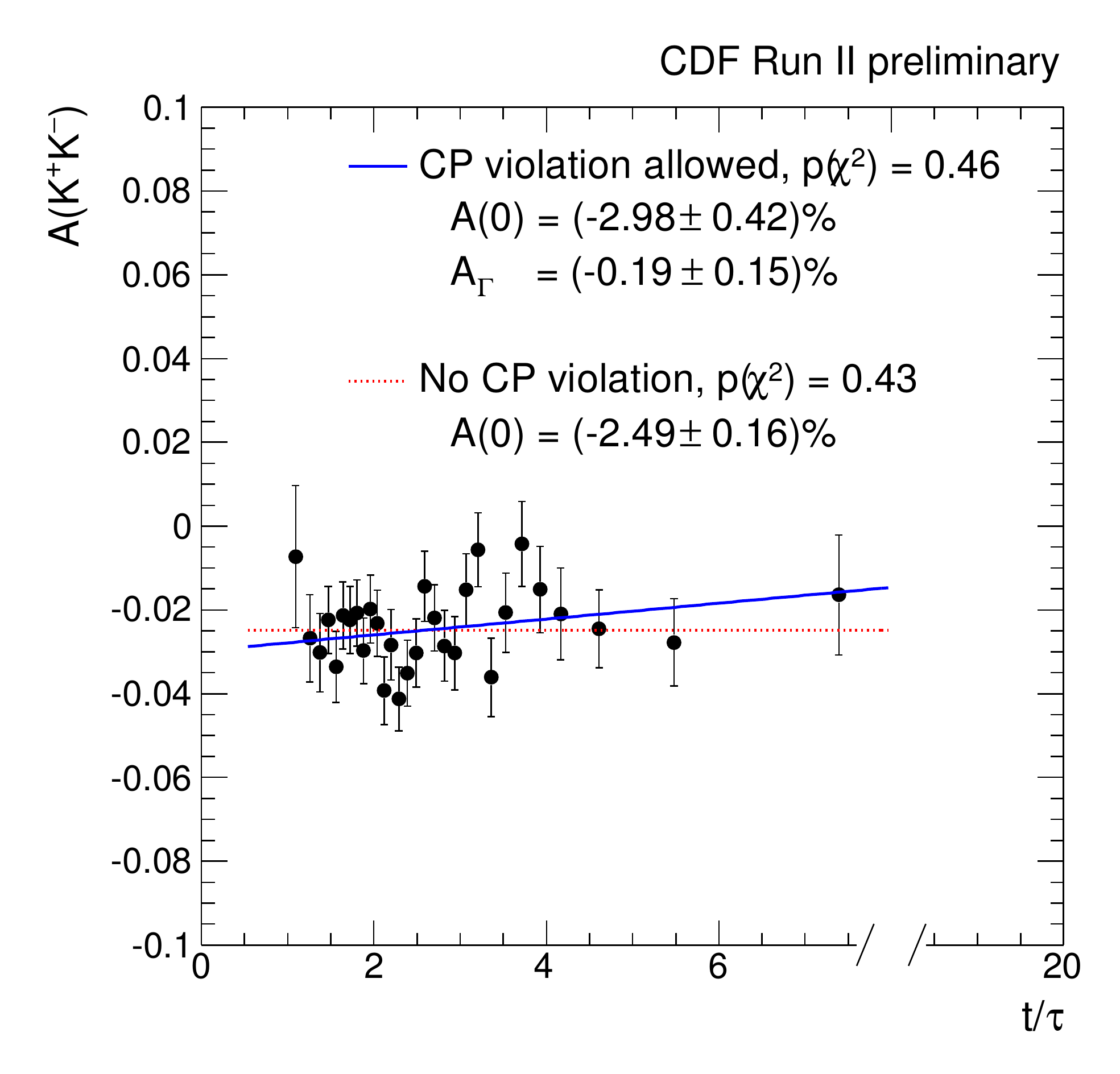}
\put(25,25){(a)}
\end{overpic}\hfil
\begin{overpic}[width=0.45\textwidth]{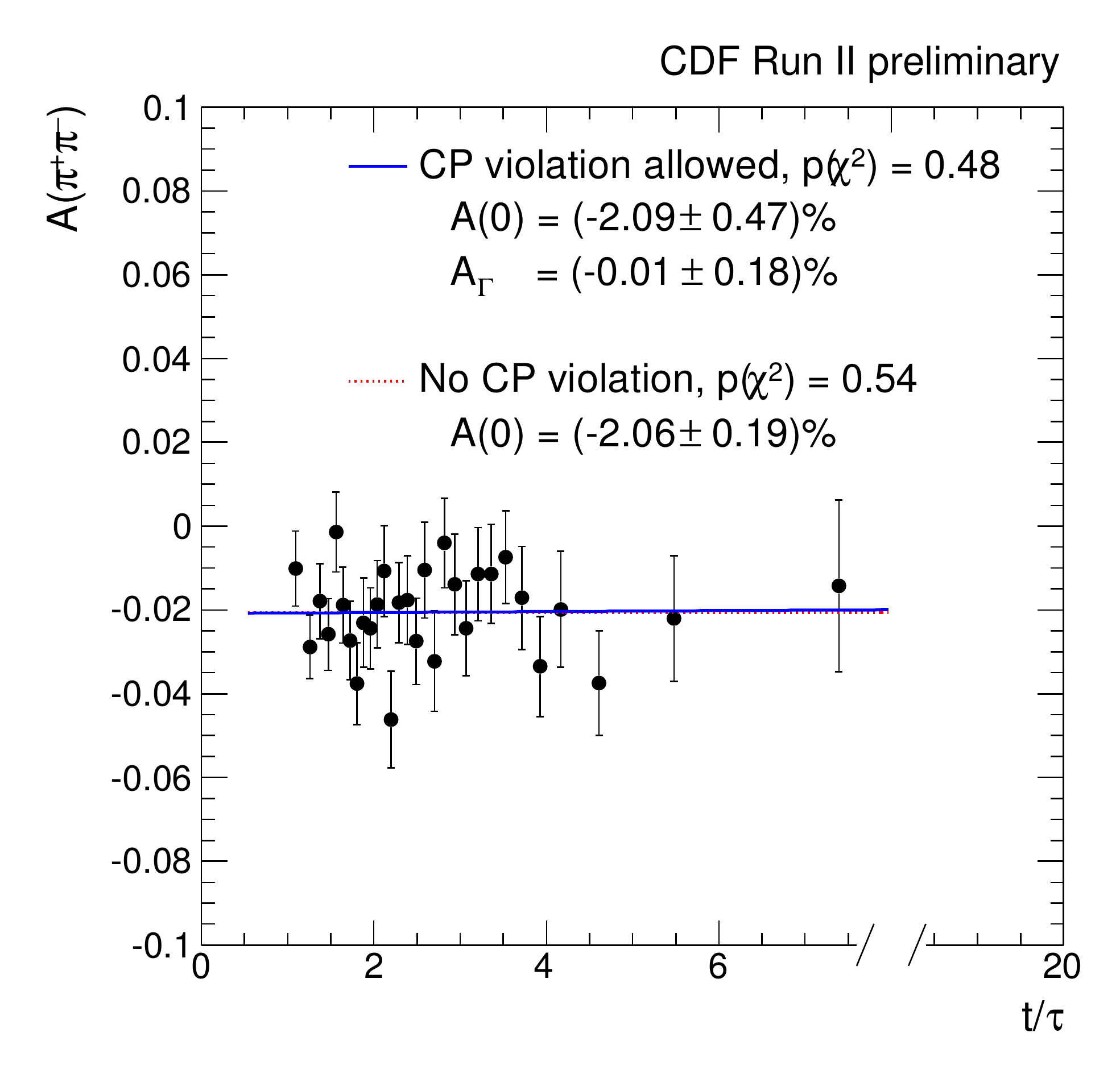}
\put(25,25){(b)}
\end{overpic}\\
\caption{Effective lifetime asymmetries as functions of decay time for the (a) $D \to K^+K^-$ and (b) $D \to\pi^+\pi^-$ samples. Results of fits not allowing for (red dotted line) and allowing for (blue solid line) \CP\ violation are overlaid.}\label{fig:Agamma}
\end{figure}

The flavor-conserving strong-interaction processes $D^{*+} \to D^0 \pi^+$ and $D^{*-} \to \Dbar^0\pi^-$ allow identification of the initial flavor through the charge of the low-momentum $\pi$ meson (soft pion, $\pi_s$).  $D^0$ or $\Dbar^0$ subsamples are thus divided in equally populated 30 bins of decay time between $0.15\tau$ and $20\tau$. 
In each bin, the average decay-time $\vev{t}$ is determined from a sample of about $13\times10^{6}$ $D^{*\pm}\to D(\to K^\mp\pi^\pm)\pi_s^\pm$ signal decays. 
Signal and background yields in the signal region are determined in each decay-time bin, and for each flavor, through $\chi^2$ fits of the $D\pi_{s}^{\pm}$ mass distribution. 
The functional form of the signal shapes is determined from simulation~\cite{AcpCDF}, with parameters tuned in the sample of $D\to K^\mp\pi^\pm$ decays, independently for each $D$ flavor and decay-time bin. The resulting signal-to-background proportions are used to construct signal-only distributions of the $D$ impact parameter (IP). In each bin and for each flavor background-subtracted IP distributions are formed by subtracting IP distributions of background candidates, sampled in the $2.015 < M(D\pi^{\pm})<2.020~\massgev$ region for the $\pi^+\pi^-$ sample, from IP distributions of signal candidates which have $M(D\pi_s^\pm)$ within $2.4~\massmev$ of the known $D^{*+}$ mass~\cite{hfag}. Contamination from multibody decays in the $K^+K^-$ sample is taken into account using as background candidates in the sideband $m_{D^0}-64~\massmev<M(K^+K^-)<m_{D^0}-40~\massmev$ and with $M(D\pi_s^\pm)$ within $2.4~\massmev$ of the known $D^{*\pm}$ mass. 
A $\chi^2$ fit of these signal-only IP distributions identifies $D^{*\pm}$ mesons from $b$-hadron decays ({\it secondary})
and determines the yields of charm ($N_{D^{0}}$) and anticharm ($N_{\overline{D}^{0}}$) mesons directly produced in the $p\bar{p}$ collision ({\it primary}). 
Double-Gaussian models are used for both the primary and secondary components. The parameters of the primary component are derived from a fit of candidates in the first decay-time bin ($t/\tau<1.18$), where any bias from the $\mathcal{O}(\%)$ secondary contamination is negligible, and fixed in all fits. 
The parameters of the secondary component are determined by the fit independently for each decay-time bin, Figure~\ref{fig:mass}.
The yields are then combined into the asymmetry $A=(N_{D^0}-N_{\Dbar^0})/(N_{D^0}+N_{\Dbar^0})$, which is fit with the linear function in Eq.\ (\ref{eq:acp3}). 
The slope of the function, which yields $A_{\Gamma}$, is extracted using a $\chi^2$ fit.
The fit is shown in Fig.~\ref{fig:Agamma} and yields $A_{\Gamma}(K^+K^-) = \bigl(-1.9 \pm 1.5\stat\bigr)\times 10^{-3}$ and $A_{\Gamma}(\pi^+\pi^-) = \bigl(-0.1 \pm 1.8\stat\bigr)\times 10^{-3}$. In both samples, we observe a few percent value for $A(0)$, due to the known detector-induced asymmetry in the soft-pion reconstruction efficiency~\cite{AcpCDF}. The independence of instrumental asymmetries from decay time is demonstrated by the analysis of $D \to K^\mp\pi^\pm$ decays, where no indirect \CP\ violation occurs and instrumental asymmetries are larger; an asymmetry compatible with zero is found, $(-0.5 \pm 0.3)\times 10^{-3}$. \par
For the $\pi^+\pi^-$ analysis, the dominant systematic uncertainty of 0.028\% arises from the choice of the impact-parameter shape of the secondary component whereas for the $K^+K^-$ sample this effect only contributes 0.013\%. The choice of the background sideband has a dominant effect in the $K^+K^-$ analysis (0.038\%) and a minor impact (0.010\%) on the $\pi^+\pi^-$ result.  

\par

\section{Conclusions}

A measurement of the difference in effective lifetime between anticharm and charm mesons reconstructed in $D^0 \to K^+K^-$ and $D^0 \to \pi^+\pi^-$ decays using the full CDF data set is reported. The final results,
\beqa
A_{\Gamma}(\pi^+\pi^-) = (-0.1 \pm 1.8 \stat  \pm 0.3 \syst)\times10^{-3},\CR
A_{\Gamma}(K^+K^-) = (-1.9 \pm 1.5 \stat  \pm 0.4 \syst)\times10^{-3},
\eeqan
are consistent with \CP\ symmetry and combined to yield $A_\Gamma = \bigl(-1.2 \pm 1.2)\times10^{-3}$~\cite{Aaltonen:2014efa}. The results are also consistent with the current best results \cite{BfactoriesAgamma}, have the second best precisions,  and contribute to improve the global constraints on indirect \CP\ violation in charm meson dynamics.  


\end{document}